\documentclass[pre,twocolumn,showpacs,aps,floats,floatfix,superscriptaddress]{revtex4}

\usepackage{subfigure}
\usepackage{amsmath, amsthm, amssymb}

\usepackage[T1]{fontenc}
\usepackage{inputenc}
\usepackage{graphicx}
\usepackage{hyperref}

\usepackage{dcolumn}
\usepackage[english]{babel}
\usepackage{times}
\usepackage{color}
\usepackage{psfrag}

\begin{document}

\title{Detecting Robust Patterns in the Spread of Epidemics: A Case Study of Influenza
in the United States and France}

\author{Pascal Cr{\'e}pey}
\affiliation{Unit{\'e} 707, Institut national de la Santé et de la Recherche médicale, Paris, France}
\affiliation{Unité mixte de Recherche en Santé 707, Universit{\'e} Pierre et Marie Curie, Facult{\'e} de M{\'e}decine Pierre et 
Marie Curie, Paris, France}
\author{Marc Barth\'elemy}
\affiliation{Centre d'Etudes de Bruy{\`e}res-le-Ch{\^a}tel, DAM Ile-de-France, Commissariat à l'énergie
atomique, Bruy\`eres-Le-Ch\^atel, France}
\affiliation{School of Informatics and Center for Biocomplexity, Indiana University, Bloomington, IN}

\date{\today}
\widetext

\begin{abstract}
In this paper, the authors develop a method of detecting correlations between epidemic patterns in different
regions that are due to human movement and introduce a null model in which the travel-induced correlations are
cancelled. They apply this method to the well-documented cases of seasonal influenza outbreaks in the United
States and France. In the United States (using data for 1972-2002), the authors observed strong short-range
correlations between several states and their immediate neighbors, as well as robust long-range spreading patterns
resulting from large domestic air-traffic flows. The stability of these results over time allowed the authors to
draw conclusions about the possible impact of travel restrictions on epidemic spread. The authors also applied this
method to the case of France (1984-2004) and found that on the regional scale, there was no transportation mode
that clearly dominated disease spread. The simplicity and robustness of this method suggest that it could be
a useful tool for detecting transmission channels in the spread of epidemics.

\end{abstract}

\maketitle


Understanding quantitatively how a disease spreads in
modern society is a crucial issue. In particular, the high
probability of occurrence of the next influenza pandemic
raises interest in the design of efficient containment policies
\cite{Ferguson:2005, Longini:2005, Ferguson:2006, Mills:2006, Colizza:2007}
and necessitates an accurate characterization of spatiotemporal
epidemic patterns. Recent outbreaks of highly
communicable diseases \cite{Lloyd:2003, Lipsitch:2003, Choi:2003, Dye:2003, Donnelly:2003, Wang:2004}
have triggered a series of
studies on the mechanisms of global disease spread \cite{Colizza:2007, Hufnagel:2004, Meyers:2005, Colizza:2006}
, and other studies have addressed the issue at the
level of individual countries \cite{Ferguson:2005, Longini:2005, Viboud:2006, Brownstein:2006}. In all of these
studies, the travel and movement of individuals is a crucial
point \cite{Colizza:2007, Hufnagel:2004, Colizza:2006, Brownstein:2006, Rvachev:1985, Flahault:1990, Flahault:1988}.
It is of the highest importance for
control strategies to identify the main channels of transmission
or "epidemic pathways", if any \cite{Colizza:2006}. Indeed, identifying
such pathways provides a first hint on how to control
a disease's spread. Even if, in most cases, travel restrictions
are economically unrealistic, knowing the most important
transmission channels could help to slow down the epidemic
through the use of selective travel restrictions.
In addition to epidemics of emergent diseases, recurrent
influenza epidemics are a burden for societies located in
temperate areas. They affect approximately 5-15 percent
of the population worldwide and are responsible for
250,000-500,000 deaths annually \cite{WHO:2007}. Different influenza
surveillance systems have been set up in various parts of the
world \cite{Valleron:1986, Bravata:2004, CDC:2007}, and influenza has been tracked for a long
time (the World Health Organization Global Influenza
Surveillance Network was established in 1952), which
makes it a well-documented disease. These data provide
an important ground for testing and developing strategies
and for uncovering the main transmission mechanisms of
a disease at different scales. Indeed, the spread of influenza
has been the focus of many studies for several years, and
a number of models have been proposed to describe and
understand it \cite{Rvachev:1985, Bonabeau:1998, Viboud:2004, Baroyan:1969, Anderson:1991}. Other investigators have thoroughly
described the spatial distribution of influenza spread
\cite{Greene:2006, Sakai:2004} and of annual waves of infection in the United
States \cite{Viboud:2006, Brownstein:2006}.
Because of the complexity of epidemic processes, together
with the "noise" present in data, innovative statistical
analyses need to be developed for detecting patterns. For
example, using signal processing methods, Brownstein et al.
\cite{Brownstein:2006} recently found evidence of correlations between domestic
airline volume and the transnational spreading time
of influenza. We think that at this point the convergence of
different methods is critical, and in this paper we propose
another method which requires little data manipulation and
filtering. In this paper, we discuss our results in the light of
recent studies \cite{Viboud:2006, Brownstein:2006} and bring our perspective to issues
such as the existence of preferred channels of transmission
and the impact of travel restrictions.
We have developed a robust and relatively simple method
of detecting correlations between different areas--"robust
spatial patterns"--in empirical data on disease dynamics. In
this paper, we illustrate and apply this method to the well documented
cases of influenza dynamics in the United
States and France. Our goal was to detect genuine correlations
between different regions and to use as few assumptions
as possible. The problem originates in the fact that
a correlation coefficient usually aggregates different phenomena.
In addition, large correlation coefficients can arise
from external large-scale environmental constraints but do
not reflect the existence of actual correlations due to other
factors such as human movements. In order to characterize
the level of correlation in a particular system under study,
we need a reference (or "null") model which gives us the
corresponding value of the correlation coefficient in the
absence of transportation flows. In this paper, we propose
a method of obtaining such a null model and apply it to the
dynamics of influenza transmission in the United States and
France. In both cases, we find robust transmission channels
for epidemic spread. In a second step, we try to relate the
existence of such channels to transportation flows.

\section{Material and Methods}
\label{sec:matmed}

\subsection{Data sets}
\label{ssec:intra-inter-fluct}
We analyzed interpandemic influenza epidemics occurring
during the periods 1972-2002 for the United States
and 1984-2004 for France. In both cases, we defined an
influenza "epidemic period" as a year running from September
to September, in order not to truncate the epidemic
season. For the United States, we used weekly state-specific
mortality rates for pneumonia and influenza collected by the
Centers for Disease Control and Prevention, restricted to the
48 contiguous states and the District of Columbia. According
to Viboud et al. \cite{Viboud:2006} and Greene et al. \cite{Greene:2006}, the weekly
time series of pneumonia and influenza mortality appear to
be useful indicators of the time evolution of spatial spread
and, interpreted cautiously, incidence within each state. For
France, we used estimates of the daily incidence of influenza-
like illness in each region, and we restricted our study
to the 21 continental regions of France. Those estimates were
based on data collected by the Sentinel Network \cite{Valleron:1986}, a network
of general practitioners distributed throughout the
entire French territory. The estimates were in agreement with
drug-sales data, which confirms their relevance \cite{Vergu:2006}.
We investigated a possible relation between our indicator
and several types of transportation flow. Data on yearly air traffic
and commuter volumes in the United States, by state,
were obtained from the Bureau of Transportation Statistics
(unpublished data; http://www.bts.gov) and the Census
Bureau \cite{Census:2007}. Data on French interregional train-traffic volume
per year were obtained from the French National Railway
Service (unpublished data; http://www.sncf.fr), and interregional
automobile traffic volume was based on 2001 data obtained
from the Service d'Etudes techniques des Routes et
Autoroutes (unpublished data; http://www.setra.equipement.
gouv.fr). We also took into account geographic distances between
states or regions, approximated by using the distances
between state capitals for the United States and the distances
between regional prefectures for France.
To investigate the possible effect of climate, we used
weekly temperature data for US states for the period 1995-2006, 
obtained from the National Weather Service \cite{NWS:2007}.

\subsection{Methods}
\label{ssec:met}
We investigated separately in the two data sets a correlation
coefficient-based indicator which enabled us to assess
the importance of travel flows. For the United States, results
were computed over 30 years of weekly data, and for France
they were computed over 20 years of daily data. For all pairs
of areas $i$ and $j$, we computed the usual Pearson correlation
coefficient $r_{ij}$ for incidence over each epidemic period (see
Appendix). The value of this coefficient is usually difficult
to interpret, however, and needs a comparison value. Thus,
we propose below a simple way to cancel correlations in real
data (uncorrelated model). We also estimate the maximal
correlation that one can observe in the system under study.
These values, obtained for the uncorrelated and maximally
correlated models, define an interval which allows for
a quantitative estimate of the correlations existing between
different areas.
\begin{figure}[t!]
  \vskip .5cm
  \begin{center}
    \includegraphics[width=8cm]{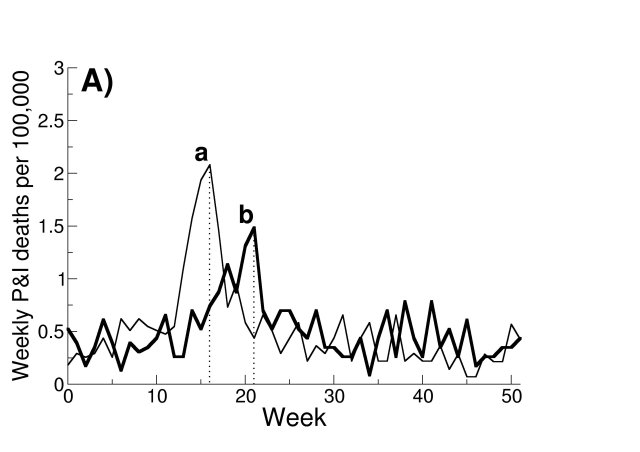}\\
    \includegraphics[width=8cm]{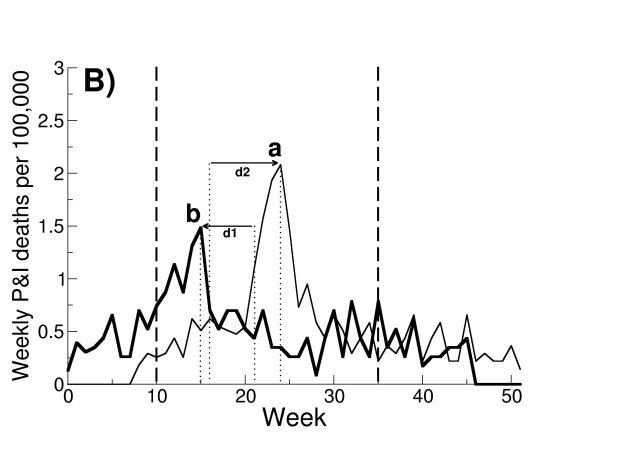}\\
    \includegraphics[width=8cm]{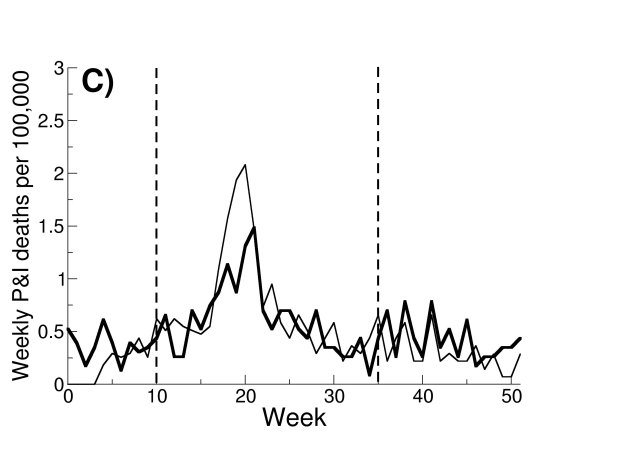}\\
    \caption{Model of maximal correlation and uncorrelated epidemic
disease spread. A) Weekly rate of mortality from pneumonia and
influenza (P\&I) per 100,000 for two different US states. The letters
a and b and the dotted lines indicate the corresponding epidemic
peaks. B) Random reassignment of the peak moments, resulting in
a shift of the different epidemic profiles by randomly chosen amounts
d1 and d2. C) The specific shift that gives the highest correlation
coefficient for the pair of epidemic peaks and corresponds to
a synchronization of the peaks. For parts B and C, the dashed lines
indicate the "epidemic activity time range."
    }
    \label{fig:un}
  \end{center}
\end{figure}

{\it Uncorrelated model.} If the spread of a disease in a country
is essentially due to the travel of infected persons between
cities, there will be a particular time ordering of the
activity profiles for different areas. The extreme case corresponds
to carriers who can make only short-range displacements,
leading to a spatial diffusion phenomenon with
a well-defined epidemic front propagating at a given velocity
(which was the case, for example, for the Black Death
\cite{Langer:1964, Noble:1974}). In this case, an outbreak appears at a time that is
directly related to the distance from the initial infectious
seed. In modern societies, people can travel over large distances,
which dramatically modifies the spatial diffusion of
the disease and its simple propagating front; however, there
should still be a pattern. Indeed, if an infected person carries
the disease from an infected area $A$ to a noninfected area $B$
and transmits the disease in area $B$, we will observe a time
difference between the epidemic peaks in the two areas. We
can expect that the larger the flow between areas $A$ and $B$,
the greater the similarity between the epidemic profiles in
the two areas. The time ordering and the correlation between
the epidemic profiles in different regions is thus a signature
of the flow of infected individuals.
Unfortunately, epidemic activities usually occur over
a short period of time in all regions of a country, and a large
observed correlation could result from this short-period constraint
without producing significant information about
transmission channels. The aim of the natural null model
is to eliminate the large correlation value occurring by
chance. Therefore, it corresponds to a random time shift
of the activity profiles of the different regions. If the values
of the actual correlation coefficient are then significantly
larger than the ones obtained with the null model corresponding
to epidemics emerging at random times, it is the
signature of genuine correlations induced by the displacement
of individuals from one region to another.
For every epidemic period (from September to September),
we define the "epidemic activity time range", which contains
all peaks for all regions or states. We then shift, by a random
amount drawn from a uniform probability, the whole epidemic
profile of a region or state, such that the epidemic profile
still belongs to the epidemic activity time range (figure \ref{fig:un},
parts $A$ and $B$). The next step is to compute the Pearson
correlation coefficients for all pairs of areas--for a given
year--and for a large number of random shifts. We finally
obtain the average correlation coefficient $m_{ij}$ between all pairs
of areas $i$ and $j$ for the uncorrelated (or "null") model.

{\it Maximal correlation.} The correlation between two areas
will be maximal for very similar "synchronized" activity
profiles with activity peaks reached essentially at the same
time. In order to obtain this maximal value for the correlation
coefficient for a given pair of areas and a given year, we
compute the cross-correlation of the epidemic profiles of the
two areas. When the maximum correlation coefficient is
found, we store it and reapply the method to the next pair
of areas. Parts $A$ and $C$ of figure \ref{fig:un} illustrate the method and
show epidemic profiles with the shift that produces the maximum
correlation coefficient (corresponding to epidemic
peak synchronization as expected). The output is a matrix
$M_{ij}$ of maximal possible correlation coefficients for the pair
of areas $(i, j)$ for a given year.

{\it A correlation coefficient-based indicator.} The uncorrelated
model gives the value of the correlation coefficient
in the absence of time correlations in the peak value of
the epidemic period, while the maximal value is obtained
when the peaks are synchronized. We can combine these
different values in order to obtain a parameter $X_{ij}(t)$ between
areas $i$ and $j$ for year $t$:
$$
X_{ij}(t) = \frac{r_{ij}(t)-m_{ij}(t)}{M_{ij}-m_{ij}(t)},
$$
where we recall that $r_{ij}$, $m_{ij}(t)$, and $M_{ij}(t)$ denote the correlation
coefficients obtained previously (calculated for year $t$).
The coefficient $X$ is thus bounded by 1 by construction, and
since time reshuffling cancels essentially time-ordering correlations,
poorly correlated and anticorrelated states can exist
and have $r_{ij} < m_{ij}$, leading to negative values of $X$. 
Those values can even be very low when $M_{ij}-m_{ij} << 1$. Below, we
analyze these quantities X for every year.

{\it Robust patterns.} We are looking for spatial patterns due
to persistent factors which do not vary significantly from
one year to another. We are thus interested in pairs of areas
showing both a high value of $X_{ij}$ and a low dispersion around
this average, which indicates a regularity in the large values
of $X$ and hence a robustness in the bond between the corresponding
areas. As we will see, most links have a relatively
large average value $\langle X \rangle$, and the low level of fluctuation is
simply characterized by the inverse coefficient of variation
($CV$) of $X_{ij}$, defined as
$$
CV(X_{ij})=\frac{\sqrt{\langle X_{ij}(t)^2\rangle - \langle X_{ij}(t)\rangle^2}}{\langle X_{ij}(t)\rangle}.
$$
A low standard deviation will then be associated with a high
value of $1/CV(X_{ij})$.

{\it Spatial autocorrelation.} In order to determine the presence
of spatial correlations in our indicator, we use Moran's
$I$, which is a weighted correlation coefficient for $X$ where the
weights depend on the distance h between two regions.
In our results we use binary weights, where 1 is attributed
to pairs of states at a distance between $h$ and $h + a$ and 0
otherwise.

\section{Results}

\begin{figure}[t!]
  \vskip .5cm
  \begin{center}
    \includegraphics[width=8cm]{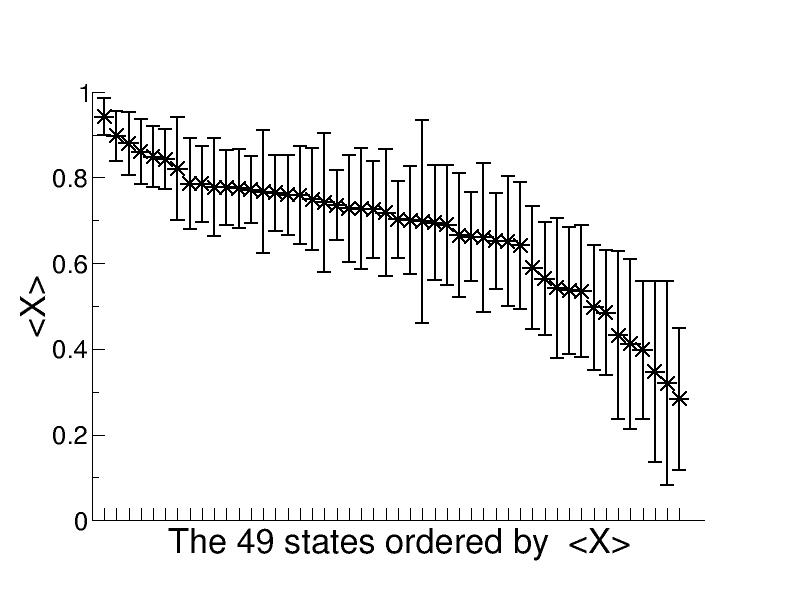}\\
    \caption{Average correlations of epidemic profiles between
California (y-axis) and the other 48 US states (47 contiguous states
plus the District of Columbia) (x-axis), computed for 30 epidemic
periods (1972--2002). Data are ranked from the highest value to the
lowest. Bars, 95\% confidence interval.
	    }
\label{fig:deux}
\end{center}
\end{figure}

In figure \ref{fig:deux}, we provide an example of average values of $X$
over the period 1972--2002 between California and other US
states. As we can see, most connections have a large average
correlation, and the main point of interest in figure \ref{fig:deux} concerns
the heterogeneity of confidence intervals, which give
information about the stability of the bond between two
areas from year to year. Indeed, one can observe that some
bonds have a highly fluctuating correlation (for Wisconsin,
95 percent confidence interval: 0.46, 0.92), while others
display a remarkable robustness (for Arizona, 95 percent
confidence interval: 0.84, 0.94), which is the signature of
a stable recurring pattern.

\begin{figure}[h!]
  \vskip .5cm
  \begin{center}
    \includegraphics[width=8cm]{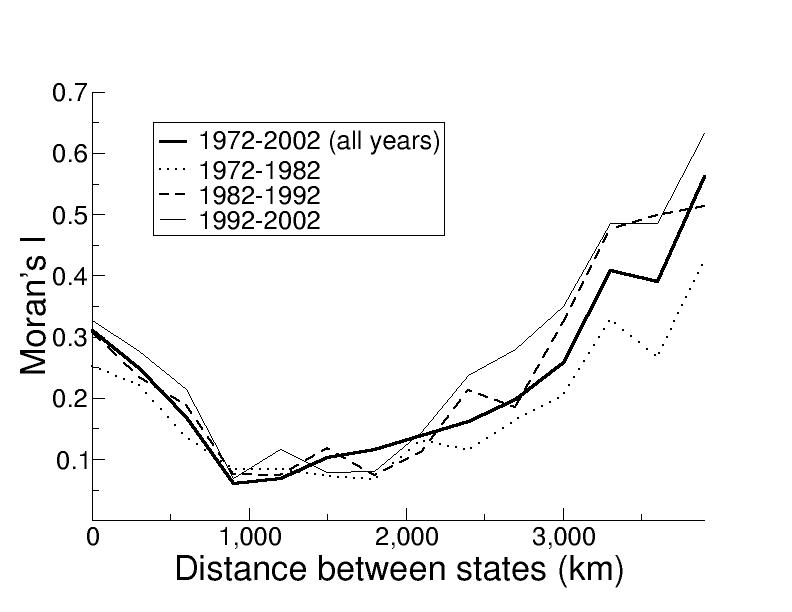}\\
    \caption{Spatial autocorrelation of X. The graph shows Moran's I
    computed over X, plotted as a function of the distance between states.
    Pairs of states are grouped by the distance between them, within
    a 300-km band. The bold line shows the evolution of the correlation
    measure for X averaged over 30 years (1972--2002). The thinner lines
    stand for X averaged over three different decades. The curves all
    display the same trends, showing stability of the spreading pattern
    over time.
    }
\label{fig:trois}
\end{center}
\end{figure}

We also must consider spatial autocorrelation analysis on
the X indicator. We aggregate data for all of the states and
show, in figure \ref{fig:trois}, the evolution of Moran?s I with different
threshold distances h. In the same plot, we show the results
for X averaged over all years in our data set (1972--2002) or
averaged over three different decades. The spatial autocorrelation
reveals correlation clusters at both short and long
distances. Figure \ref{fig:trois} also shows that our results are stable over
time.

\begin{figure}[h!]
  \vskip .5cm
  \begin{center}
    \includegraphics[width=8cm]{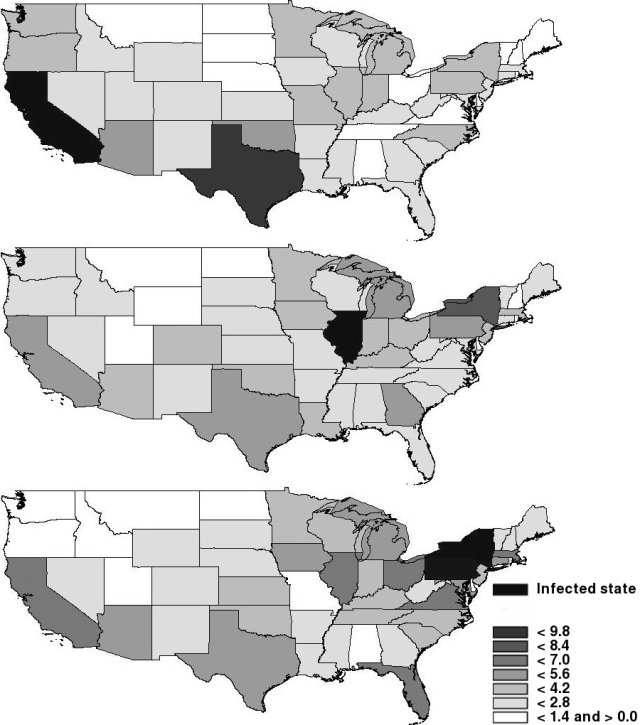}\\
    \caption{Correlations of epidemic profiles for three US states
    (denoted as "infected state"). The various shades of gray stand for
    values of $1/CV(X_{ij})$, where $i$ corresponds to California, Illinois, or New
    York (top to bottom) and $j$ to all of the other states. In each map, black
    stands for the considered state (California, Illinois, or New York). We
    observe high values of $1/CV(X_{ij})$ for neighboring states (Arizona for
    California; Indiana, Kentucky, and Ohio for Illinois; New Jersey,
    Pennsylvania, and Massachusetts for New York) and long-range
    connections between California and Texas, Illinois and New York, and
    New York and California.
    }
\label{fig:quatre}
\end{center}
\end{figure}

\begin{figure*}[th]
  \vskip .5cm
  \begin{center}
    \includegraphics[width=0.9\textwidth, height=4cm]{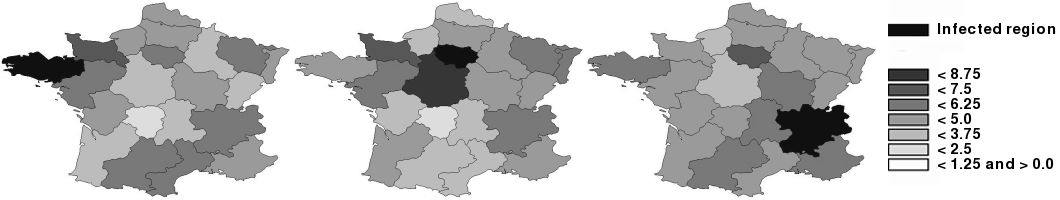}
    \caption{Correlations of epidemic profiles for three regions in France (denoted as "infected region"). The various shades of gray stand for
    values of $1/CV(X_{ij})$, where $i$ corresponds to Bretagne, Ile-de-France, or Rhone-Alpes (left to right) and $j$ to all of the other regions. In each map, black
    stands for the considered region (Bretagne, Ile-de-France, or Rhone-Alpes).
    }
\label{fig:cinq}
\end{center}
\end{figure*}

Figure \ref{fig:quatre} displays the results for the correlation ($1/CV(X_{ij})$)
on maps for three different states. In the three situations,
we observe high $1/CV(X_{ij})$ values for neighboring
states, as well as long-range bonds. This method can also be
used on the smaller scale of French regions.We also observe
two types of strong connections here (figure \ref{fig:cinq}). Regions are
strongly correlated with their neighbors, and we observe long-range
strong connections, as in the case of the United States.

\subsection{Relation with transportation data}
\begin{figure}[h]
  \vskip .5cm
  \begin{center}
    \includegraphics[width=8cm]{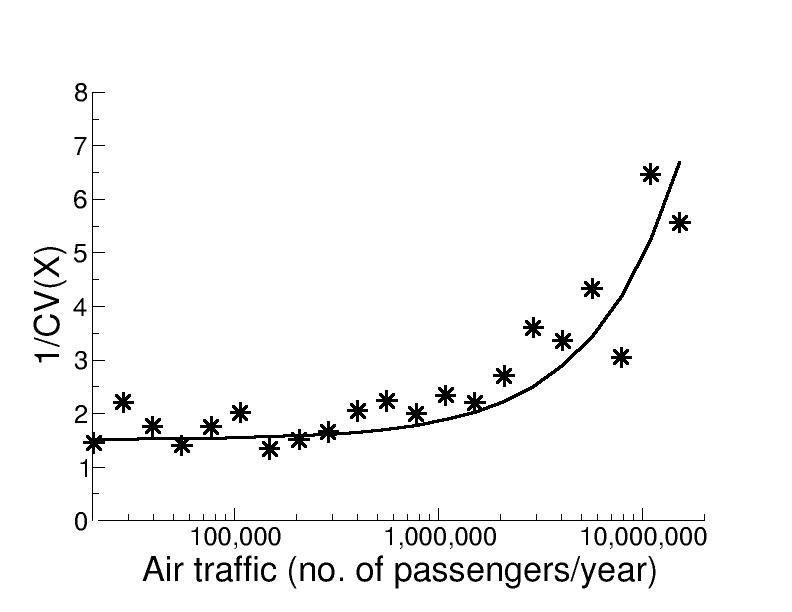}\\
    \caption{Correlation indicator for air transportation flow in the
    United States. The graph shows the inverse coefficient of variation
    ($CV$) of X as a function of domestic air transportation flow for all pairs
    of states (binned data). Since the range of variation of the traffic was
    very wide, the log-linear plot is shown. Air transportation was measured
    as the number of passengers traveling by plane between pairs
    of states in 2000. The curve is a linear fit of the equation $1/CV(X) = A * flow + B$, where $A = 2.5 * 10^{-7}$ and $B = 0.91$ ($r^2 = 0.74$).
    }
\label{fig:six}
\end{center}
\end{figure}

\begin{figure}[h]
  \vskip .5cm
  \begin{center}
    \includegraphics[width=8cm]{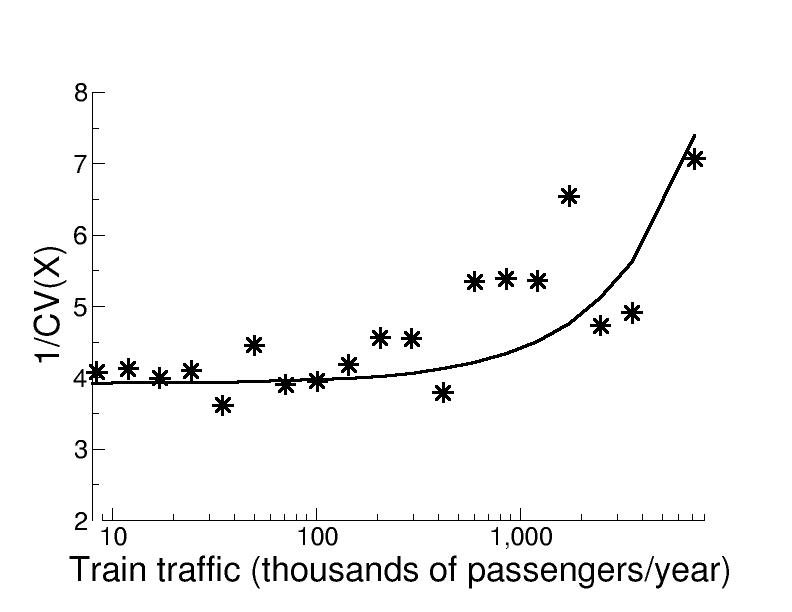}\\
    \caption{Correlation indicator for train traffic in France. The graph
    shows the inverse coefficient of variation (CV) of X as a function of
    train transportation for all pairs of regions (binned data). Since the
    range of variation of the traffic was very wide, the log-linear plot is
    shown. Train transportation was measured as thousands of passengers
    traveling by train between pairs of regions in 2001. The curve is
    a linear fit of the equation $1/CV(X) =  A * flow + B$, where
    $A = 4.6 * 10^{-5}$, $B = 3.9$ ($r^2 = 0.52$).
    }
\label{fig:sept}
\end{center}
\end{figure}

In figures \ref{fig:six} and \ref{fig:sept}, we plot the quantity $1/CV(X)$ against
transportation traffic data to test its relation to transportation
flows. In figure \ref{fig:six}, we plot $1/CV(X)$ against the interstate US
air-traffic flow. The linear fit computed has a coefficient of
determination of 0.74 and thus supports evidence that $1/
CV(X)$ is proportional to the air-traffic flow (note that we
have not normalized the traffic with respect to the population).
This plot supports the claim that $1/CV(X)$ indicates
that the main vector for the spread of an epidemic in the
United States is domestic air traffic.
In order to assess the specific contribution to our indicator
of air traffic as compared with other parameters, such as
temperature or geographic distance, we performed a multivariate
regression analysis using a linear model. For
temperature, we used the Pearson correlations of weekly
reported temperatures between states.We used state temperature
profiles covering 11 years to control for possible artifacts.
The results of this analysis (table \ref{table:un}) suggest that
interstate air traffic makes a greater contribution than distance
or temperature (estimates were 0.407, 0.096, and
0.220, respectively). Although the model does not explain
the total variance ($r^2 = 0.27$), it is statistically significant
($p < 0.001$) and suggests that air travel is the dominant factor
among those factors explored.

As recent studies have shown \cite{Viboud:2006, Riley:2006}, travel to work \cite{Census:2007}
plays a predominant role in disease spread, and thus we
estimated its impact on our results. The volume of interstate
commuting is very low (representing approximately 3 percent
of the total number of US commuters) and decreases
quickly with distance, so we needed to consider only pairs of
states with common borders. Through a multivariate regression
on $1/CV(X)$, we compared the impacts on epidemic
spread of air travel and commuting using ground modes of
transportation. Air traffic made a larger contribution in the
regression model (table \ref{table:deux}), and we can thus conclude that it
is the dominant transportation mode, even between neighboring
states.
On a smaller scale, the results for France were more contrasted.
As figure \ref{fig:sept} shows, the behavior of $1/CV(X)$ increased
with train traffic, but we also performed a linear
multivariate regression analysis of $1/CV(X)$ as a function
of train traffic, automobile traffic, and geographic distance.
Our aim was not to obtain a fully explicative model but to
test whether one of these factors would dominate the others.
The results (table \ref{table:trois}) show that distance makes little contribution
to the model and that automobile and train traffic
have essentially the same weights (0.092,
0.265, and 0.204, respectively). As in the US case, the
model does not explain all of the variance ($r^2 = 0.22$), but
the result is statistically significant ($p < 0.001$).

\begin{table}[h]
\begin{tabular}{p{0.35\textwidth}p{0.10\textwidth}}
\toprule
	& Estimate\\
\hline
Intercept & 0\\
A: Air traffic volume & 0.407\\
B: Distance between states & -0.096*\\
C: Temperature correlation & 0.220*\\
\hline
\end{tabular}\\
\vskip 0.2cm
$* p < 0.001$. 
\label{table:un}
\caption{The linear regression equation takes the form $X=b_1+b_2*A+b_3*B+b_4*C$,
where $A$, $B$, and $C$ stand for standardized air traffic volume, distances between
states, and correlation of state temperatures, respectively, and $b_j$ are the estimates
given in the table ($r^2=0.27, p<0.001$).}
\end{table}

\section{Discussion}
We have presented a method that aims to identify strongly
connected areas from raw epidemic data. Our main goal is to
identify preferred spatial paths--or epidemic pathways--
for the spread of infectious diseases. If these paths exist,
they have an effect on the spatiotemporal pattern of reported
cases of disease; consequently, we should be able to identify
them from the temporal evolution of local influenza incidence.
An important fact is the need to discriminate what is
due to inherent noise, spatial effects, and other constraints in
the data. In order to achieve this goal, we define and use an
uncorrelated model which cancels the existing correlations
due to transportation flows. The maximal correlation gives
the upper bound of the possible correlation that would exist
between two areas if they were perfectly synchronized.
These lower and upper bounds enable us to assess the level
of genuine correlation due to transportation flow between
regions. By observing data from several years, we can analyze
the evolution of the level of correlation between areas
and detect the robustness of patterns over time.The spatial autocorrelation analysis showed that the behavior
of our indicator is stable over time. Results were
consistent from one set of years to another, despite possible
environmental evolution. This stability, particularly with respect
to the increase in air-travel flow (on the order of 300
percent between 1972 and 2002), seems to indicate that for
some time now, air-travel flows have been large enough to
propagate an epidemic throughout the United States. This
implies that in order to be efficient, travel restrictions should
be so drastic that they are economically unreasonable--
a finding that agrees with other recent results \cite{Colizza:2007, Viboud:2006}. The
spatial regression analysis also revealed the existence of
geographic clusters of strongly correlated neighbor states
at short distances (<600 km) and other clusters at long
distances (>3,000 km). We cannot relate these results directly
to the smaller scale (the intercounty level) studied by
Viboud et al. \cite{Viboud:2006}, since we were using aggregated data, but
as expected, our results were in agreement with those of that
study at the larger interstate scale.

\begin{table}[h]
\begin{tabular}{p{0.35\textwidth}p{0.10\textwidth}}
\toprule
	& Estimate\\
\hline
Intercept & 0\\
A: Air traffic & 0.341\\
B: Ground commuting & 0.068\\
\hline
\end{tabular}\\
\vskip 0.2cm
$* p < 0.001$. 
\label{table:deux}
\caption{The linear regression equation takes the form $X=b_1+b_2*A+b_3*B+b_4*C$,
where $A$, $B$, and $C$ stand for air traffic and commuting between
states, respectively, and $b_j$ are the estimates
given in the table ($r^2=0.12, p<0.001$).}
\end{table}

The next step is to interpret these patterns in terms of
social, geographic, or other epidemiologic data. In the case
of the United States, we were able to relate the existence of
persistent channels of transmission to interstate air transportation
flows, while other factors such as climate seemed
less important. We also showed that even between neighboring
states, air travel is dominant over commuting using
ground modes of transportation. Commuters very likely
play a role in the spread of epidemics at the county level,
and further investigation at this smaller scale is needed.
More generally, such an interpretation might not be particularly
feasible because of the mixing of different modes
of transportation. Indeed, the case of France, which is of
interest for its smaller scale, highlights the weakness of
a hypothesis pertaining to a single mode of transportation.
An explanation for this might be that different modes of
travel (train, automobile, and plane) compete at this scale
and there is no clearly dominant transportation mode, a result
which is supported by our multivariate analysis. The
existence of a large $1/CV(X)$, however, reveals the existence
of strongly connected regions. The lack of clear correlations
between large values of $1/CV(X)$ and large traffic volumes
does not affect the quality of the indicator $1/CV(X)$ but
simply reflects a more mixed situation concerning transportation
modes used in the influenza epidemic process.
There are different limitations to this work. In particular,
pneumonia and influenza mortality (for the United States)
or influenza-like illness (for France) are just proxies for
laboratory-confirmed influenza. Moreover, we did not
take into account the change in influenza strains from
one year to another. However, these factors are unlikely to
have affected our conclusions about robust transmission
channels. Another possibly important limitation concerns
the fact that all studies (including ours) are limited to the
spread of disease inside a given country and neglect the
exchange of disease with other countries. The corresponding
flows are usually far from negligible, and we think their
importance in national spread should be assessed in future
studies.
\begin{table}[h]
\begin{tabular}{p{0.35\textwidth}p{0.10\textwidth}}
\toprule
	& Estimate\\
\hline
Intercept & 0\\
A: Distance between regions & -0.092$\dag$\\
B: Automobile traffic volume & 0.265*\\
C: Train traffic volume & 0.204*\\
\hline
\end{tabular}\\
\vskip 0.2cm
$* p < 0.001$, $\dag p<0.1$. 
\label{table:trois}
\caption{The linear regression equation takes the form $X=b_1+b_2*A+b_3*B+b_4*C$,
where $A$, $B$, and $C$ stand for standardized, distances between
regions, automobile traffic volume, and train traffic volume, respectively, and $b_j$ are the estimates
given in the table ($r^2=0.22, p<0.001$).}
\end{table}
Our heavy use of retrospective data to compensate for
data fluctuations did not allow us to analyze and interpret
single-year fluctuation as shown in the paper by Brownstein
et al. \cite{Brownstein:2006}. Basically, those authors chose a different tradeoff
than ours and decided to aggregate their data in a few US
geographic regions, whereas we decided to keep a statewise
approach and use as many years as possible.
Our results seem to suggest that for the United States,
realistic modeling of the spread of epidemics at the interstate
level may only need to take air transportation into
account. They also seem to imply that in the case of France,
a realistic model would need to include several transportation
modes.
In summary, we believe that this simple and robust
method, which detects important channels of disease transmission,
could be helpful in modeling the spread of epidemics
and in assessing containment strategies that rely on travel
restrictions.


\begin{acknowledgments}
P. C. received financial support from Action Concert\'ee Incitative 
"Syst\`emes Complexes en Sciences Humaines et Sociales."
The authors thank Dr. C\'ecile Viboud for interesting suggestions 
and discussions and for making her data available. 
M. B. thanks the School of Informatics at Indiana University, where this work was started.
Conflict of interest: none declared.
\end{acknowledgments}


\appendix
    \section*{Appendix}
For all pairs of areas i and j, we compute the usual Pearson
correlation coefficient for disease incidence over each
epidemic period (composed of n time steps), as
$$
r_{ij}=\frac{\sum_{t=1}^n{(x_i(t)-\rangle x_i\langle)(x_j(t)-\langle x_j\rangle)}}{nS_iS_j}
$$
where $x_i(t)$ and $x_j(t)$ are the estimated incidences for areas $i$
and $j$ at time step $t$, $\langle x_i\rangle$ and $\langle x_j\rangle$ are their averages, and $S_i$ and
$S_j$ are their standard deviations.
\end{document}